\documentstyle[aps,epsfig,multicol]{revtex}

\def\beq{\begin{equation}}
\def\eeq{\end{equation}}

\begin{document}

\title{Particles Sliding on a Fluctuating Surface: Phase Separation and Power Laws} 
\author{Dibyendu Das and Mustansir Barma}
\address{Department of Theoretical Physics, Tata Institute of Fundamental Research, Homi Bhabha Road, Mumbai 400 005, India}
\maketitle

\begin{abstract}
We study a system of hard-core particles sliding downwards on a fluctuating 
one-dimensional surface which is characterized by 
a dynamical exponent $z$. In numerical simulations, an initially random 
particle density is found to coarsen and obey scaling with a growing length 
scale $\sim t^{1/z}$. The structure factor deviates from the Porod law
in some cases. The steady state is unusual in that the
density-segregation order parameter shows strong 
fluctuations.  The two-point correlation function
has a scaling form with a cusp at small argument which we relate
to a power law distribution of particle cluster sizes. 
Exact results on a related  model of surface depths provides insight 
into the origin of this behaviour.


\end{abstract}
\begin{multicols}{2}

How do density fluctuations evolve in a system of particles moving on
a fluctuating surface?  Can the combination of random vibrations and an
external force such as gravity drive the system towards a state with
large-scale clustering of particles?  Such large-scale clustering driven
by a fluctuating potential represents an especially interesting 
possibility for the behaviour of two coupled systems, one of which evolves
autonomously but influences the dynamics of the other. 
Semi-autonomous systems are currently of interest in diverse contexts, 
for instance, advection of a passive scalar by a fluid \cite{Kraichnan}, 
phase ordering in rough films \cite{Drossel}, the motion of stuck and 
flowing grains in a sandpile \cite{Biswas}, and 
the threshold of an instability in a sedimenting colloidal
crystal \cite{Lahiri}.

In this paper, we show that there is an unusual sort of phase 
ordering in a  simple model of this sort, namely a
system of particles sliding downwards under a gravitational field 
on a fluctuating one-dimensional
surface. The surface evolves through its own dynamics, while the
motion of particles is guided by local downward slopes; since random 
surface vibrations cause slope changes, they constitute a source of
nonequilibrium noise for the particle system.  
The mechanism which promotes clustering is simple: fluctuations lead 
particles into potential minima or valleys, and once together the particles 
tend to stay together as illustrated in Fig. 1. The question is whether this 
tendency towards clustering persists up to macroscopic scales. We show
below that in fact the particle density exhibits coarsening towards a 
phase-ordered state. This state has uncommonly large fluctuations which 
affect its properties in a qualitative way, and make it quite 
different from that in other driven, conserved, systems  which exhibit 
coarsening \cite{Cornell}.

It is useful to state our principal results at the outset. (1) In an infinite
system, an initially randomly distributed particle density exhibits coarsening
with a characteristic growing length scale ${\cal{L}}(t) \sim t^{1/z}$ where
$z$ is the dynamical exponent governing fluctuations of the surface. 
For some of the models we study, the scaled structure factor varies 
as $|{k{\cal{L}}(t)}|^{-(1+\alpha)}$ with $\alpha < 1$,
which represents a marked deviation from the Porod law ($\alpha = 1$) for 
coarsening systems \cite{Bray}. Further, a finite system of size $L$ reaches
a steady state with the following characteristics: (2) The magnitude
of the density-segregation order parameter has a nonzero time-averaged
value, but shows strong fluctuations which do not decrease as $L$ increases.
(3) The static two-point correlation function $C(r)$ has a cusp at
small values of the scaled separation $|r/L|$. 
(4) The sizes $l$ of particle clusters are distributed according to a
power law for large $l$, up to an $L$-dependent cutoff.
These results are established by extensive numerical simulations. 
Further, the properties (3) and (4) are shown to be related,
using the independent interval approximation \cite{Satya} applied 
to the cluster distribution.  Also,
we define a related coarse-grained depth model of the surface, and show
analytically that the steady state characteristics (2)-(4) hold for this
model.

\begin{figure}[tb]
\begin{center}
\leavevmode
\epsfig{figure=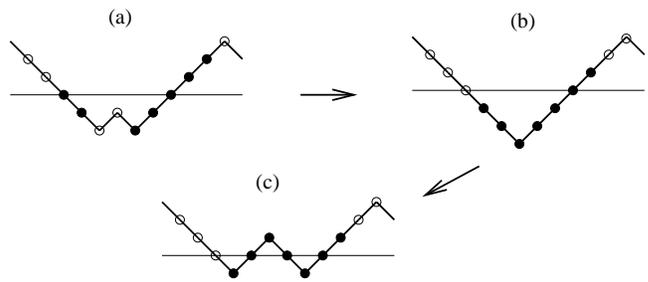,width=8.5cm}
\end{center}
\narrowtext
\noindent\caption{Depicting clustering of 
particles ($\bullet$) in a section of the fluctuating surface. 
A surface fluctuation 
($a$) $\rightarrow$ ($b$) causes the particles to roll into a valley. 
Particles remain clustered even after a reverse surface 
fluctuation ($b$) $\rightarrow$ ($c$) occurs.}
\label{fig:land}
\end{figure}

The Sliding Particle (SP) model is defined 
as a lattice model of particles moving on a 
fluctuating surface. Both the particles and the surface degrees of freedom
are represented by $\pm 1$-valued Ising variables $\{\sigma_{i}\}$ and 
$\{\tau_{i-{1 \over 2}}\}$ 
on a one-dimensional lattice with periodic boundary conditions, where 
$\sigma$ spins occupy lattice sites, and ${\tau}$ spins the links between 
sites. Then $n_{i} = {1 \over 2}(1 + {\sigma_{i}})$ represents the particle 
occupation of site $i$, whereas 
$\tau_{i-{1 \over 2}} = +1$ or $-1$ represents the 
local slope of the surface (denoted $/$ or $\backslash$, respectively). The 
dynamics of the interface is that of the single-step model {\cite{Plischke}},
with stochastic corner flips involving exchange of adjacent $\tau$'s;
thus, $/ \backslash \rightarrow \backslash /$ with rate $p_{1}$, while 
$\backslash / \rightarrow / \backslash$ with rate $q_{1}$. A particle and a 
hole on adjacent sites ($i$,$i+1$) exchange with rates that depend on the 
intervening local slope $\tau_{i-{1 \over 2}}$;  thus the moves $\bullet 
\backslash \circ \rightarrow \circ \backslash \bullet$ and $\circ / \bullet
\rightarrow \bullet / \circ$ occur at rate $p_{2}$, while the inverse moves
occur with rate $q_{2} \neq p_{2}$. The asymmetry of the rates reflects the 
fact that it is easier to move downwards along the 
gravitational field. Note that the dynamics conserves 
$\sum \sigma$ and $\sum \tau$; we work in the sector where both vanish. 
In the remainder of the paper,
we report results for symmetric surface fluctuations ($p_{1} = q_{1}$), 
whose behavior at large length and time scale is described by the continuum
Edwards-Wilkinson model \cite{Edwards}. 
Further we consider the strong-field ($q_2 = 0$) limit for the
particle system, and set $p_2 = p_1$.
We have also investigated the behaviour away from these limits,  
and found that our broad conclusions remain unaffected. The SP
model is a limiting case of the Lahiri-Ramaswamy (LR) model of
sedimenting colloidal 
crystals {\cite{Lahiri}}; it corresponds to the tilt field 
evolving autonomously. As we shall see below, this change causes 
strong phase separation {\cite{Lahiri}} to disappear, and a new 
type of phase ordering, characterized by strong fluctuations, to appear in 
its stead. 
 
In the SP model, particles preferentially occupy the lower portions or
large valleys of the fluctuating surface.
In order to study the dynamics of the hills and valleys
of the surface, we define a height profile $\{ h_i \}$ 
with $h_i = {\sum_{1{\leq}j{\leq}i}} \tau_{j-{1 \over 2}}$. 
We then define a Coarse-grained 
Depth (CD) model by considering spins $s_{i} = - sgn(h_i)$ where $s_i$  
is $+1$, $-1$ or $0$ if 
the surface height $h_i$ at site $i$ is below, above or at the zero 
level. 
A stretch of like $s_i$'s $= +1$ represents a valley with respect to the 
zero level. The time evolution of the CD model variables $\{ s_i \}$ is
induced by the underlying dynamics of the bond variables 
$\{ \tau_{i-{1 \over 2}} \}$. The model is similar to the domain growth 
model of Kim {\it et al} {\cite{Kim}}.

We studied the evolution of the density in the SP model starting from an 
initial random placement of particles on the fluctuating 
surface. After an initial quick downward slide into local 
valleys, the density distribution is guided by the 
evolution of the surface profile.  
To quantify the tendency towards clustering, we monitored the 
equal time correlation function ${\cal C}(r,t) \equiv \langle \sigma_{i}(t) 
\sigma_{i+r}(t) \rangle$ 
by numerical simulation (Fig. 2). If $z$ is the dynamical 
exponent characteristic of the surface fluctuations ($z = 2$ for the 
symmetric surface model), 
we expect the scale ${\cal L}(t)$ for density fluctuations to be set by the 
base lengths of typical coarse-grained hills which have overturned in time $t$,
i.e. ${\cal L}(t) \sim t^{1/z}$. This is indeed the case in the scaling
limit ($r >> 1$, $t >> 1$, $r/{{\cal L}(t)}$ fixed) as shown by the 
collapse to a scaling function ${\cal C}_s(y = r/{{\cal L}(t)})$ 
in Fig. 2. We have also checked that similar scaling collapses occur
when we use other models of surface fluctuations with widely different values
of $z$ ($z = 3/2$ for Kardar-Parisi-Zhang (KPZ) {\cite{KPZ}} 
surfaces, and $z \simeq 4$ for the Das Sarma-Tamborenea model {\cite{DT}}). 

The existence of a single growing length scale ${\cal L}(t)$ is indicative of 
coarsening towards a phase ordered state {\cite{Bray}}. In the SP model, 
coarsening is driven by surface fluctuations, rather than more customary
temperature quenches. This causes an interesting feature to appear in the 
scaling function ${\cal C}_s$, namely a distinctive cusp for small argument:  
${\cal C}_s(y) = {\cal C}_0 - {\cal C}_1 {y}^{\alpha}$ for $y << 1$ (Fig. 2).  
We find the cusp exponent $\alpha \simeq 0.5$. 
This cusp implies that the scaled structure 
factor ${\cal S} \sim (k{\cal L})^{-{(1+\alpha)}}$ for large $k{\cal L}$.
This is substantially different from the Porod law behavior 
$(k{\cal L})^{-2}$, characteristic of customary coarsening 
systems {\cite{Bray}}. 

\begin{figure}[tb]
\begin{center}
\leavevmode
\epsfig{figure=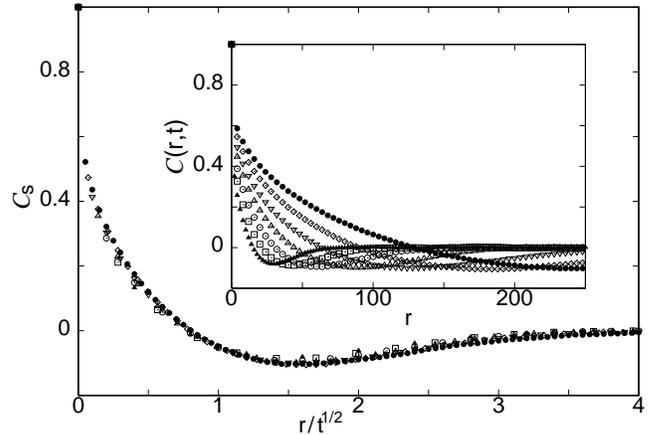,angle=-90,width=8.5cm}
\end{center}
\narrowtext
\noindent\caption{The data for ${\cal C}(r,t)$
at different times $t = 400{\times}2^n$ (with $n = 1$,...,$6$) shown in the
inset, is shown to collapse when scaled by ${\cal L}(t) \sim t^{1/2}$.}
\label{fig:ewh}
\end{figure}
 
In the steady state, in order to characterize phase ordering  of the SP model,
we monitored the magnitude of the Fourier components of the density profile
\begin{equation}
Q(k) = |{{1 \over L} {\sum_{j=1}^L} e^{ikj} n_j}|,~~~~~~~~~~ 
k = {2{\pi}m \over L}.
\label{qk} 
\end{equation}
where $m = 1,...,L-1$. 
As in {\cite{Korniss}} we used the lowest nonzero Fourier component $Q^* 
\equiv Q({2\pi \over L})$ as a measure of the phase separation in our
system with conserved dynamics. The time average $\langle Q^{*} \rangle \simeq
0$ in a disordered state,
and is $\simeq 0.318$ in a fully phase separated state. For the 
SP model, Fig. 3 
shows numerical results for $\langle Q(k) \rangle$ versus $k$ for various 
system sizes. While $\langle Q^* \rangle$ approaches a finite limit as 
$L \rightarrow \infty$, the values of Fourier components at fixed $k$ decrease
with increasing $L$. This provides strong evidence for phase separation,  
corresponding to the occurrence of 
density inhomogeneities of the order of the system size. 
However $Q^*(t)$ shows strong fluctuations as a function of time (Fig. 3, 
inset). With increasing $L$, the separations between fluctuations 
increase with $L$, but the amplitude of fluctuations does not decrease. 
The best way to 
characterize these strong fluctuations is through the full probability
distribution $Prob(Q^*)$ which we found numerically. 
The mean value $\langle Q^* \rangle \simeq 
0.18$, while the RMS fluctuation 
$(\langle {Q^*}^2 \rangle - {\langle Q^* \rangle}^2)^{1/2} 
\simeq 0.07$. Despite these large fluctuations in $Q^*$, the configuration of 
the system does not become randomly disordered even when $Q^*(t)$ is small. 
Rather, we found that the next few Fourier
modes ($k = 4\pi/L$, $6\pi/L$, $...$) are excited at such times, 
indicating a break-up into a few macroscopic regions of different density.

\begin{figure}[tb]
\begin{center}
\leavevmode
\epsfig{figure=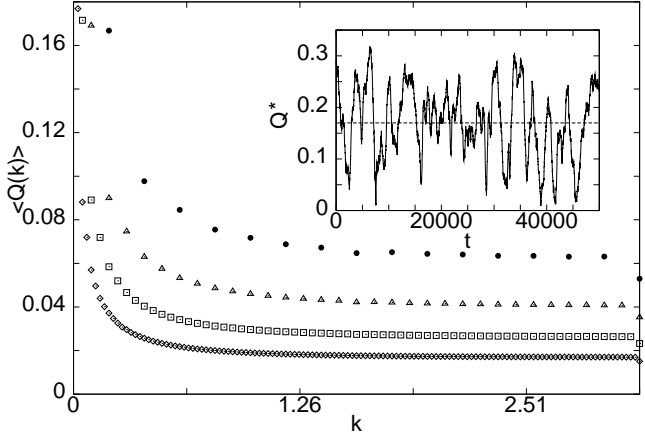,angle=-90,width=8.5cm}
\end{center}
\narrowtext
\noindent\caption{$\langle Q(k) \rangle$ as a function of $k$, for 
different sizes $L = 32, 64, 128$ and $256$ (from above to below). 
Inset: Time variation of $Q^*$ in the steady state for $L = 128$.} 
\label{fig:Q(k)}
\end{figure}

Not surprisingly, these macroscopic fluctuations leave a strong imprint on 
spatial correlation functions and cluster distributions. For instance, 
the two-point correlation function $C(r) \equiv \langle \sigma_j \sigma_{j+r}
\rangle$ varies with $r$ on the scale of the system size $L$, as is evident 
from Fig. 4 which shows that the data collapse onto a single 
curve when plotted versus $r/L$.
For comparison recall that a phase-separated system with sharp interfaces 
between two macroscopic phases would show a linear decrease $C(r) = 
{M_{o}^{2}}(1 - 2|r|/L)$ on length scales larger than the correlation length.
By contrast, the curve in Fig. 4 is nonlinear in the 
full range of $r/L$. Further as with the coarsening correlation function, the
scaling function $C_s$ for steady state correlations has a cusp for small
argument:
\begin{equation}
C_s({r \over L}) = c_0 - c_1{|{r \over L}|}^{\alpha},~~~~~~~~~
|{r \over L}| << 1
\label{scale}
\end{equation}
with $\alpha \simeq 0.5$. We will see below that 
this is related to the form of the size distribution of clusters, defined as
groups of contiguous lattice sites occupied by particles. Figure 5 shows
that this distribution follows a power law $P(l) \sim l^{-\theta}$ with 
$\theta \simeq 1.8$.

\begin{figure}[tb]
\begin{center}
\leavevmode
\epsfig{figure=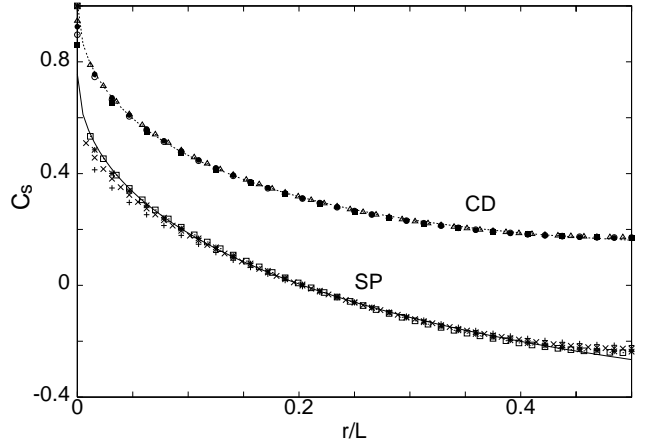,angle=-90,width=8.5cm}
\end{center}
\narrowtext
\noindent\caption{$C(r)$ in the steady state of the SP and the CD 
models for $L = 64$, $128$, $256$, and $512$. Notice the cusp at small 
$|r|/L$. The curves are fits to the form $c_o - c_1 y^{1/2} + c_2 y$.}
\label{fig:steady}
\end{figure}

\begin{figure}[tb]
\begin{center}
\leavevmode
\epsfig{figure=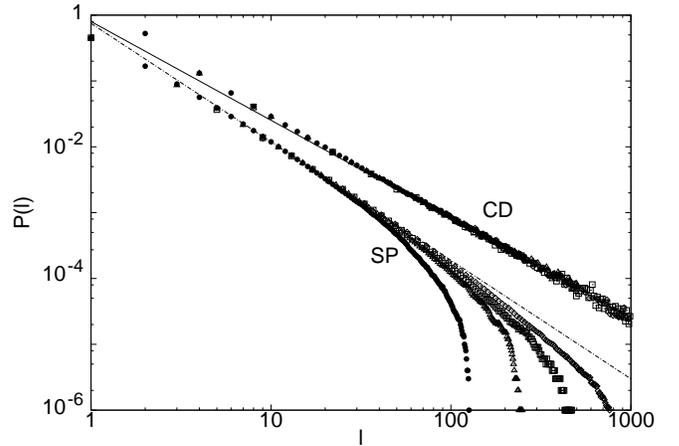,angle=-90,width=8.5cm}
\end{center}
\narrowtext
\noindent\caption{Probability distribution $P(l)$ of particle cluster lengths 
for the SP and CD models for $L = 256, 512, 1024$ and $2048$.   
The straight lines have slopes ${-1.8}$ and ${-1.5}$, respectively.}
\label{fig:P(l)}
\end{figure}

The relationship between $C(r)$ and $P(l)$ may be understood within the 
independent interval approximation (IIA) 
{\cite{Satya}}. Within this scheme, the 
joint probability of having $n$ successive intervals is treated as the 
product of the distribution of single intervals. In our case, the intervals 
are successive clusters of particles and holes, which occur with probability
$P(l)$. Defining the Laplace transform ${\tilde P}(s) = {\int_{0}^{\infty}}
dl e^{-ls} P(l)$, and ${\tilde C}(s)$ analogously, we have {\cite{Satya}}
\begin{equation}
s(1 - s{\tilde C}(s)) = {2 \over {\langle l \rangle}} 
{{1 - {\tilde P}(s)} \over {1 + {\tilde P}(s)}}
\label{iia1}
\end{equation}
where $\langle l \rangle$ is the mean cluster size. In usual applications of 
the IIA, the interval distribution $P(l)$ has a finite first moment 
${\langle l \rangle}$ independent of $L$. But that is not the case here, as 
$P(l)$ decays as a slow power law $P(l) \sim l^{-{\theta}}$ for $l >> 1$. 
Since the largest possible value of $l$ is $L$, we have ${\langle l \rangle} 
\approx a L^{2-\theta}$ for large enough $L$. Considering $s$ in the range 
$1/L << s << 1$, we may expand ${\tilde P}(s) \approx 1 - b s^{\theta - 1}$;  
then to leading order, the right hand side of Eq. ({\ref{iia1}}) becomes 
$bs^{\theta - 1}/aL^{2-\theta}$, implying ${\tilde C}(s) \approx 
1/s - b/(a L^{2-\theta} s^{3 - \theta})$. This leads to 
\begin{equation}
C(r) \approx 1 - {b \over {a \Gamma(3 - \theta)}}|{r \over L}|^{2 - \theta}.
\label{iia2}
\end{equation}
This has the same scaling form as Eq. ({\ref{scale}}).
Matching the cusp singularity in Eqs. (\ref{scale}) and (\ref{iia2}), we get 
\begin{equation}
\theta + \alpha = 2~~~~~~~~~~~~~~~~~~~~~~~~~~~~~(\rm IIA).
\label{iia3}
\end{equation}
By comparison, the numerically determined values for the SP model 
yield $\theta + \alpha \simeq 2.3$. 
We conclude that the IIA provides a useful insight into the 
relationship between cluster statistics and correlation functions,
even though it is not exact.

We can gain considerable understanding into the nature of phase ordering in the
SP model by analyzing the closely related CD model of surface
fluctuations. The relationship between these models, which is plausible
at a qualitative level, can be quantified by checking whether the
overlap $O= \langle \sigma_i s_i \rangle$ is nonzero; numerically 
we find $O \simeq 0.26$. The similarity
in the behaviour of the two models is brought out in Figs. 4 and 5;
$C(r/L)$ has a cusp, and $P(l)$ a power law decay, for both models.

We now show that the CD model can be mapped onto a random walk (RW) problem 
and that the mapping can be exploited to 
give exact results for several properties
of the model. We make a correspondence between each surface configurations 
and an RW trajectory, by interpreting $\tau_{i-1/2}=+1$
or $-1$ as the rightward or leftward RW step at the $i$'th time instant. Then
$s_i=1,-1$ or 0 depending on whether the walker is to the right, to the left
or at the origin after the $i$'th step. Evidently, the lengths of clusters
of $s=1$ spins (or $s=-1$ spins) represent times between successive
returns to the origin. Thus, $P(l)$ for the CD model is just the well-known
distribution for RW return times to the origin, 
which behaves as $l^{-3/2}$ for large $l$. Thus $\theta = 3/2$ in this model.

Since successive RW returns to the origin are independent events, the IIA is 
exact for the CD model. Thus Eq. (\ref{iia3}) holds, and we conclude that 
the correlation-function cusp exponent $\alpha =1/2$ for this model.

In systems with strong macroscopic fluctuations such as the SP and CD models,
characterization of the phase separated steady state requires the full 
probability distribution of the order parameter. For the CD model, an 
appropriate (nonconserved) order parameter is $M = {1 \over L}\sum {s_i}$, 
which for the RW represents the excess time a walker spends on one side of the 
origin over the other side. In order
to respect periodic boundary conditions, we need to restrict the ensemble of
RWs to those which return to the origin after $L$ steps. The full probability
distribution of $M$ over this ensemble is known from the equidistribution 
theorem on sojourn times of a RW \cite{Feller}: $Prob(M) = 1/2$ for 
$M \in [-1$,$1]$, i.e. every allowed value of $M$ is equally likely. This 
implies $|{\langle M \rangle}| = 1/2$ and $(\langle M^2 \rangle - {\langle 
|M| \rangle}^2)^{1/2} = 1/{\sqrt 12}$. The strong fluctuations in $M$ mirror 
the large fluctuations of $Q^*$ in the SP model.

To summarize, both the SP and CD models exhibit a phase ordered steady state 
with unusual fluctuation characteristics, manifested in several related ways:
slow power law decays of the cluster size distribution, a cusp singularity in
the scaled two-point correlation function, and a probability distribution 
for the order parameter which remains broad even in the thermodynamic limit.

We have checked numerically that this sort of state survives
in the SP model even when we depart from the strong field limit by allowing 
$q_2$ to be nonzero, and when we vary the ratio $p_2/p_1$ of rates of particle
and surface motion from high to low values. Finally, we found that
even when we allow different rates $p_1$ and $q_1$ for upward and downward 
corner flips which makes surface fluctuations KPZ-like, this type of phase 
separation persists and the scaled correlation function shows a cusp 
\cite{Das}. This could have interesting implications for the behaviour of 
domain walls in the problem of phase ordering on a rough surface 
\cite{Drossel}. The state is destroyed, however, if we allow for surface 
kinematic waves, which transport transverse fluctuations of the interface 
at a finite speed through the system.
It would be interesting to investigate the long-time 
dynamics of the large-scale fluctuations in the models discussed above, 
and to characterize the steady state in higher dimensions.

We acknowledge very useful discussions with M.R. Evans, D. Dhar, S. Ramaswamy, 
and C. Dasgupta. We are especially indebted to S.N. Majumdar for invaluable 
suggestions and criticisms at all stages of this work.

\end{multicols}

\end{document}